\documentclass[aps,pra,a4paper,reprint,showpacs]{revtex4-1}
\usepackage{amsmath}
\usepackage{amssymb}
\usepackage{color}
\usepackage{graphicx}
\usepackage[utf8x]{inputenc}
\usepackage[T1]{fontenc}
\usepackage{ae,aecompl}
\usepackage[english]{babel}
\usepackage{hyperref}
\newlength{\figa}
\newcommand{\op}{\left \langle y \right \rangle}  
\newcommand{\ac}{\alpha_{\rm c}}        
\newcommand{\dy}{\delta y}          
\newcommand{\ro}{\left \langle \rho \right \rangle }
\newcommand{\too}{\text{--}}

\begin{document}

\title{Dimensional transitions in small Yukawa clusters}

\author{A. Radzvilavi\v{c}ius}
\affiliation{Department of Theoretical Physics, Vilnius University, 
Saul\.{e}tekio al.~9-III, LT-10222 Vilnius, Lithuania}
\author{O. Rancova}
\affiliation{Department of Theoretical Physics, Vilnius University, 
Saul\.{e}tekio al.~9-III, LT-10222 Vilnius, Lithuania}
\author{E. Anisimovas}
\affiliation{Department of Theoretical Physics, Vilnius University, 
Saul\.{e}tekio al.~9-III, LT-10222 Vilnius, Lithuania}
\date{\today}

\begin{abstract}
We provide the detailed analysis of structural transitions leading to the
rapid changes in dimensionality of small Yukawa clusters. These
transformations are induced by the variations in the shape of confinement
as well as the screening strength. We show, that even in the most primitive
systems composed of only a few strongly interacting particles, the order
parameter exhibits a power-law behavior in the vicinity of the critical
point of the continuous transition. The critical exponent $\gamma=1/2$ is
found to be universal in all studied cases, which is consistent with the
general theory of continuous phase transitions. 

\end{abstract}
\maketitle

\section {Introduction}

The emergence of ordered patterns within complex systems of interacting entities 
draws attention of researchers in diverse fields, including physics, biology, 
mathematics and computer science \cite{Koch, Wolfram, Chakravarti}. 
Confined Wigner crystals are among the most primitive systems where the phenomenon
of self-organization is observed. A small number of charged particles is 
placed into a confining potential well, where in the limit of low temperatures 
ordered structures are formed. This crystalline type of matter was successfully 
realized experimentally in a variety of well known systems, such as electrons on 
the surface of liquid helium \cite{Ikegami, Grimes}, cooled ions in traps \cite{PP} or 
strongly coupled particle clusters in complex plasmas \cite{Bonitz}.

Systems of strongly coupled particles are of high scientific interest due to
various collective phenomena, e.g. cooperative dynamics, waves and phase 
transitions. First order transitions between solid and liquid states of matter, 
namely melting and crystallization, were widely investigated in the studies 
of many-particle dusty plasma crystals, also known as Yukawa clusters 
\cite{Schella, Nosenko}. Crystalline structures formed by dust particles  
in complex plasmas turned out to be an extremely handy tool for these 
studies, as the convenient length and time scales, stability and transparency 
of these systems allow for direct optical observation and accurate measurements 
\cite{Bonitz}. On the other hand, there is another type of transitions, 
occurring in the simplest few-particle systems confined by asymmetric traps. 
These are observed when a small change in one of the control parameters causes 
a sudden change in the dimensionality of the system, and therefore are called 
dimensional or zigzag phase transitions \cite{Melzer}. Note that 
these transitions take place in finite systems and are analogous to the second 
order phase transitions commonly defined and studied in the thermodynamic limit. 

Dimensional transitions in small two-dimensional Yukawa clusters have been 
extensively studied in Ref.~\onlinecite{Sheridan} both experimentally 
and numerically. The authors demonstrated an excellent agreement between the
computed configurations of particles and the arrangements observed in complex 
plasma experiments. Structural zigzag transitions were induced by variations in 
the number of particles, the value of the Yukawa shielding parameter $\kappa$ or 
the shape of the confinement potential. A power-law behavior of the order 
parameter in the vicinity of phase transition was observed, and interpreted
as a characteristic feature of second order phase transitions.
However, the numerical values of critical exponents provided in
\cite{Sheridan} cast some doubt. These values are distinctly different from
the classical mean-field value $1/2$ which one would naturally expect in a 
finite system of just a few particles. Therefore, one of the main motivations behind 
the current work is to provide the results of numerical modelling with higher 
precision and show the universality of critical exponents. We also extend 
the investigation to the clusters of different sizes and dimensional 
transitions of other types. 

Our paper is organized as follows. In section \ref{NP} the model system is 
described and the procedure of our calculations is presented. Section \ref{Res} 
presents the results of the numerical modelling of dimensional transitions, 
grouped by their type in subsections. Main points of the article are summarized 
in section \ref{Sum}. Additionally, Appendix \ref{appendix} provides details
on the analytically solvable cases and exact values of critical parameters.

\section {Numerical procedure}
\label{NP}

We investigate numerically systems of $N$ identical particles of mass $m$ and 
charge $Q$, interacting through the Yukawa inter-particle potential. The 
interaction energy of two charges embedded in a screening environment thus reads
\begin{equation}
  V_{ij}=\frac{Q^2}{4\pi\epsilon_0}\frac{{\rm e}^{-\kappa r_{ij}}}{r_{ij}},
\end{equation}
where $\kappa$ stands for the shielding strength (inverse screening length) and 
controls the range of interaction. Particles are kept together by a harmonic 
confinement potential  
$V_{\rm c}(\mathbf r)=\frac{1}{2} m \omega_0^2 
\left( {x}^2 + \alpha^2 {y}^2 + {z}^2\right )$ in 3D or 
$V_{\rm c}(\mathbf r)=\frac{1}{2} m \omega_0^2 
\left( {x}^2 + \alpha^2 {y}^2\right )$ in 2D. 
The anisotropy parameter $\alpha$ controls the shape of the confinement, which 
reduces to the symmetric spherical or circular form at $\alpha = 1$, and takes 
the shape of oblate ($\alpha>1$) or prolate ($\alpha<1$) spheroid in 3D and
an ellipse in 2D.

In the regime of strong correlations the potential energy dominates over the 
kinetic one, and the total energy of the model system is given by
\begin{equation}
\label{eq:Ud}
  U(\mathbf{r}_1, \ldots, \mathbf{r}_N)
  = \sum_{i=1}^N \frac{1}{2} 
  \left( {x}_i^2 + \alpha^2 {y}_i^2 + {z}_i^2\right )
  + \sum_{i>j}^N \frac{1}{r_{ij}} {\rm e}^{-\kappa r_{ij}}.
\end{equation}
The units of length and energy are conveniently chosen as 
$r_0=({Q^2}/{4 \pi \epsilon_0 m\omega_0^2})^{1/3}$ and $E_0={Q^2}/{4 
\pi \epsilon_0 r_0}$. Obviously, $\kappa$ is now measured in $r_0^{-1}$. In 
two dimensions, the $z$ coordinates of all particles are set to zero, so that 
all the particles lie within the $(xy)$ plane.

Stable arrangements of particles correspond to the local and global minima of 
the potential energy (\ref{eq:Ud}). In our present work, stationary states are 
located by the method of multiple heating-relaxation cycles, based on the Monte 
Carlo and numerical minimization algorithms. The method was already proven to 
be efficient and reliable in our previous studies \cite{MusuSeni}. As it turned 
out, the potential energy landscape of (\ref{eq:Ud}) is rather complex even 
for small values of $N$ and might be described as a collection of local minima, 
separated by potential barriers of various heights. In a first stage of the 
algorithm, thermalization takes place, that is, the system is heated to the 
temperature high enough to overcome all the potential barriers. This stage is 
accomplished by performing a large number (few thousand) of Metropolis Monte 
Carlo steps \cite{Metropolis}, which leads to a configuration, that may be 
regarded as drawn randomly from the Boltzmann distribution corresponding to 
the given temperature. Each minimum controls a certain area of coordinate space, 
called basin of attraction. The area of attraction varies from basin to basin,
which means that different stable states are realized with different 
probabilities \cite{Kahlert, MusuSeni}. Some of the minima are located 
within the regions with steep walls, while others lie in broad shallow valleys 
and therefore require a considerable effort to find. In a second stage of the 
computational procedure, the temperature of the system is suddenly set to zero 
and the closest local minimum of potential energy is located by employing the
steepest descent and Newton optimization techniques. As frequently there is 
more than one local minimum \cite{MusuSeni}, the whole cycle of thermalization 
and relaxation is repeated a large number of times, to ensure that all of the 
basins are visited and all stationary points of (\ref{eq:Ud}) are revealed.

Departure of the anisotropy parameter $\alpha$ from unity breaks spherical (or 
circular in 2D) symmetry and, as a result, prolate or flattened structures 
are formed. Eventually, at the critical value of parameter $\ac$, dimensional 
transitions are observed, three-dimensional clusters become planar, 
while two-dimensional structures are transformed into linear ones.
In order to determine the critical values of $\alpha$  with high precision, 
we repeat our calculations by incrementing $\alpha$ in small steps. As it will 
become evident shortly, properties of Yukawa clusters, including critical 
values of the anisotropy 
parameter and critical exponents, depend strongly on the screening 
parameter $\kappa$. Therefore, in most cases we use four different  
values of Yukawa screening 
strength, $\kappa = 0,\, 1,\, 2,\, 3$. In the simplest case of 
$\kappa=0$, inter-particle potential reduces to the simple unscreened Coulomb 
interaction.

A second order phase transition is marked by a sudden appearance or disappearance
of some property of the system, called an order parameter, in response to a small 
change in a control parameter. We investigate dimensional phase transitions by 
keeping an eye on the total potential energy and order parameter $\op$, which 
is defined as the root mean square of the coordinate $y$:
\begin{equation}
\op = \left(\frac{1}{N} \sum_{i=1}^N y_i^2 \right )^{1/2}.
\end{equation}
Naturally, dimensional transitions are signified by a sudden change of 
$\op$ to zero.  In particular, $\op$ is a good choice for an 
order parameter, since $\op=0$ in the 1D (2D) configuration and 
$\op>0$ in the 2D (3D)  configuration. The
state variables that determine the system configuration are
then $N$, $\kappa$ and $\alpha$.

\section{Results}
\label{Res}

\subsection{2D $\rightarrow$ 1D transitions}
\label{sec:2D1D}

\begin{figure}[ht]
\centering
\includegraphics[width=\figa]{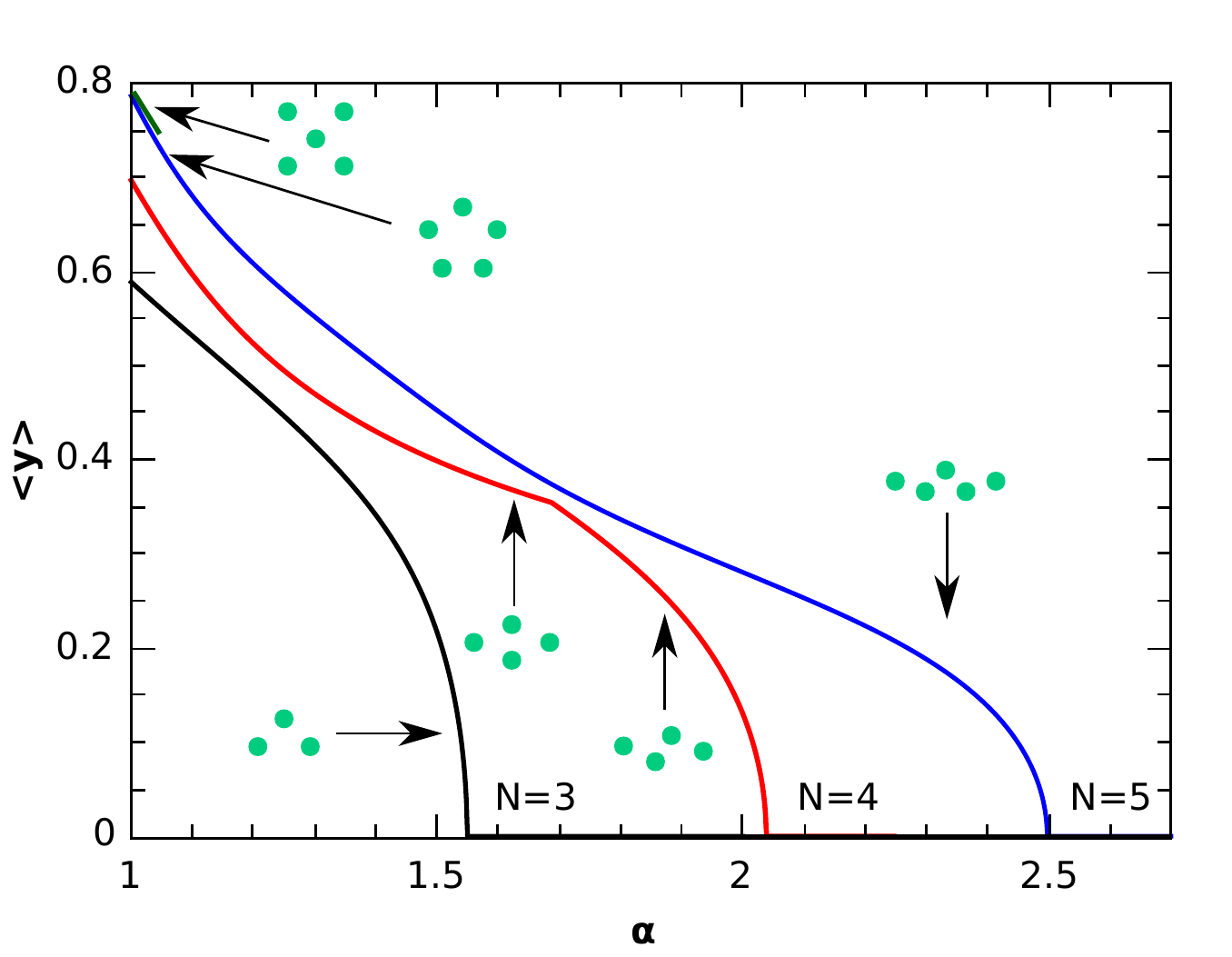}
\caption{Value of the order parameter $\op$ as a function of the 
anisotropy $\alpha$ for 2D clusters with $N=3\too 5$ particles. Insets 
show typical arrangements of particles in various stages of 
compression. Note how quickly metastable 
state $(1,\,4)$ disappears (top left corner).}
\label{fig1}
\end{figure}

We first investigate the simplest few-particle two-dimensional 
systems, undergoing $\rm 2D\rightarrow1D$ structural transitions. In 
the case of $\alpha=1$, the confinement potential is symmetric and 
particles form ordered states, that were previously observed experimentally and 
modelled theoretically \cite{Kong, Saint-Jean}. Systems with $N=3,\, 4,\, 7$ particles 
form only one stable configuration (ground state), while clusters with 
$N=5,\,6$ particles in symmetric traps support both ground- and one metastable 
configurations. Various states can be represented by listing the 
occupation numbers of different shells --- ground state of 5-particle 
system is therefore the configuration $(0,\,5)$ and metastable state is 
$(1,\,4)$ (for the arrangements of particles see Figure \ref{fig1}). As 
the anisotropy parameter $\alpha$ departs from unity, 
metastable states can become ground states, some states can 
disappear and new ones appear. In all investigated cases, however, 
there is only one stationary configuration near the dimensional 
transition --- a zigzag shaped pattern, which soon becomes a 1D linear chain 
of particles at $\alpha>\ac$.  

In the simplest symmetric case of $N=3$, particles form an 
equilateral triangle in the $(xy)$ plane. As the value of $\alpha$ 
increases, the triangular configuration is gradually deformed until 
the transition occurs at $\alpha=\ac \approx1.55$, as shown in 
Figure \ref{fig1} with $\kappa=0$. In fact, dimensional transition in 
three-particle system can be modelled analytically, which gives the value of 
$\ac=\sqrt{12/5}$ (see Appendix \ref{appendix}). Numerical simulation gives exactly the same value.

The structural transition is slightly more intriguing in the case of 
$N=4$ particles. As it can be deduced from the evolution of $\op$ in Figure \ref{fig1}, 
there is a discontinuity in a first derivative of the order parameter 
${\rm d} \op/{\rm d}\alpha$ at the value of anisotropy parameter 
$\alpha \approx 1.69$. As it turns out, there are two stages of the
four-particle cluster compression. In the first, slow stage, four particles form a 
rhombus-shaped structure, with the particles located exactly on 
$x$ or $y$ axis. Later, the stage of rapid compression takes over, 
with two particles departing from the line $x=0$ and forming a 
zigzag shaped pattern. Two typical rhombus- and zigzag-shaped configurations are presented 
in the insets of Figure \ref{fig1}. As it was shown in the previous 
study, the transition between these two 
stages is followed by the specific oscillation of the heat capacity 
\cite{Rancova}. Analogous scenario applies to the 
other clusters with even number of particles. The dimensional transition is observed at 
$\ac \approx 2.04$, where $\op$ suddenly drops to zero.

Two competing configurations are first observed in the case of $N=5$ 
particles, namely states $(0,\, 5)$ and $(1,\,4)$. As it is depicted 
in
Figure \ref{fig1}, metastable state  
$(1,\,4)$ exists only in the narrow window of anisotropy, $1<\alpha<1.05$. On the 
other hand, the pentagonal ground state undergoes a continuous 
structural transformation, forms a zigzag-shaped cluster and finally 
becomes linear at $\ac\approx2.50$.  

Structural transitions become more complex for the 
systems with $N \geq 6$. Six 
particles in a symmetric confinement can form two stable states. As 
$\alpha$ increases, metastable $(0,\, 6)$ state vanishes at 
$\alpha=1.05$ only to reappear  again
and become a new ground state later. The former ground state  $(1,\, 5)$ then 
disappears completely near $\alpha=1.22$. Six- and eight-particle 
clusters both feature the same discontinuity in ${\rm d} \op/{\rm d}\alpha$ 
as four-particle system, discussed above.

\begin{figure}[ht]
\centering
\includegraphics[width=\figa]{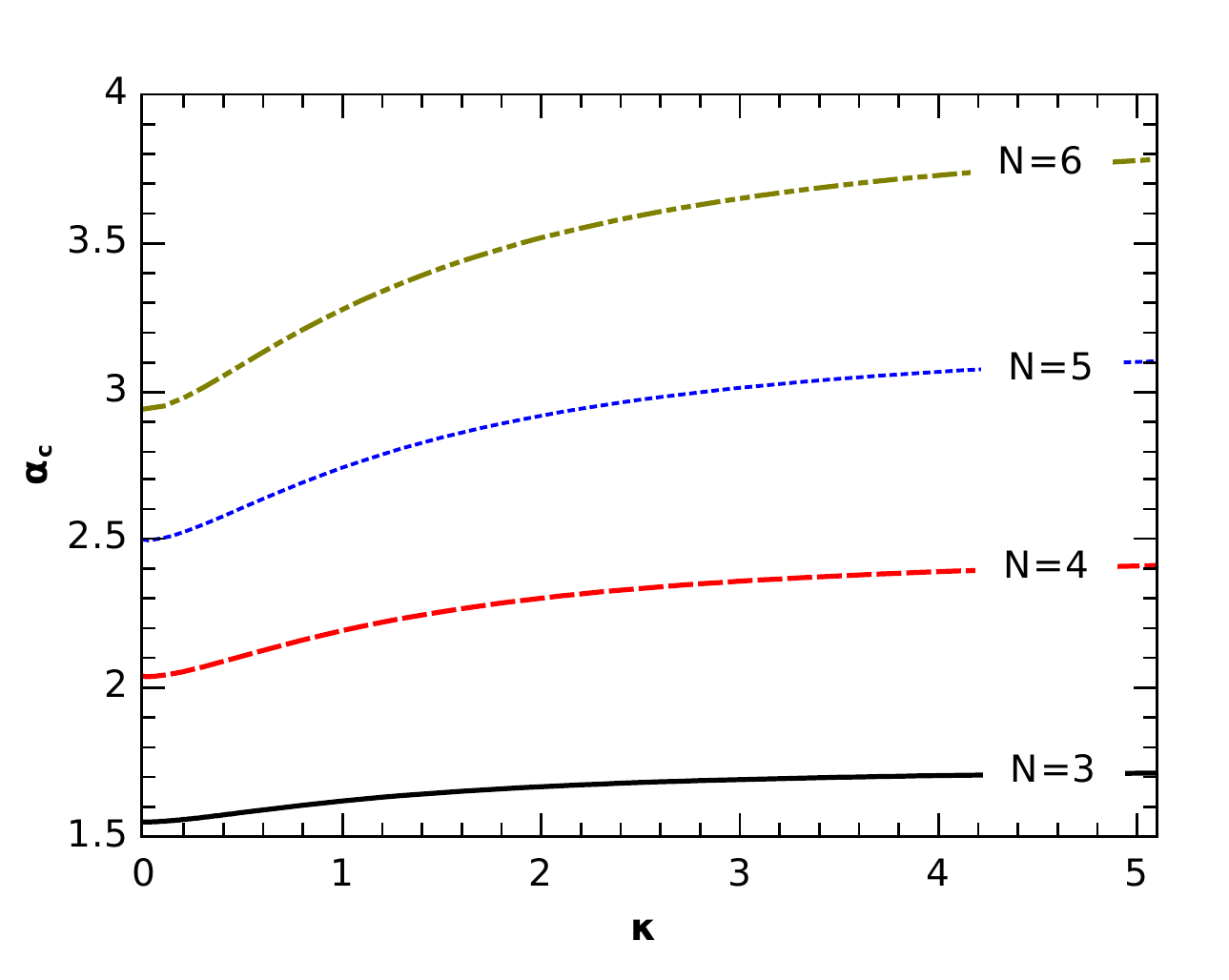}
\caption{Phase diagram for two-dimensional Yukawa clusters of $N$ 
particles. Below the corresponding line, the cluster of $N$ particles 
is two-dimensional and one-dimensional above. We see, that dimensional 
transitions can be induced by variations in any of three parameters 
$\alpha$, $\kappa$ or $N$.}
\label{fig2}
\end{figure}

We have already seen in  Figure \ref{fig1}, that critical value of 
the parameter $\ac$ increases with $N$, when inter-particle 
interaction is of the Coulomb type. As it might be expected, $\ac$ also 
grows as Yukawa potential screening parameter $\kappa$ is 
increased, which is shown in Figure \ref{fig2} for $N=3 \too 6$. The 
critical value of the anisotropy increases rapidly for 
$\kappa<1.5$ and almost saturates for high values of screening, i.e. 
$\kappa>4.0$.

The lines in Figure \ref{fig2} represent boundaries between different 
phases of clusters. The structures are two-dimensional below the line 
and form linear configurations above. Although we devote most 
of the present work to the transitions induced by deformations 
of the confinement well, structural changes can actually be caused by variations 
in any of three parameters $\alpha$, $\kappa$ or $N$. As the figure shows, two-dimensional 
cluster can become linear without any changes in $\alpha$, for 
example, when 
a value of $\kappa$ is  
diminished, or when a particle is removed from the system.

We further examine the power-law behavior of the order parameter $\op$ near 
its critical point $\ac$ in more detail. The power law is easily identified by
plotting the logarithm of the order parameter, $\lg \left( \op \right)$, as a 
function of $\lg(\ac-\alpha)$. The function turns out to be linear for small 
values of $(\ac-\alpha)$. This observation confirms that in the vicinity of 
the transition point, the order parameter $\op$ demonstrates a power-law 
behavior, which is a typical property of second-order phase transitions:
\begin{equation}
\op \propto (\ac -\alpha)^\gamma.
\end{equation}

\begin{figure}[ht]
\centering
\includegraphics[width=\figa]{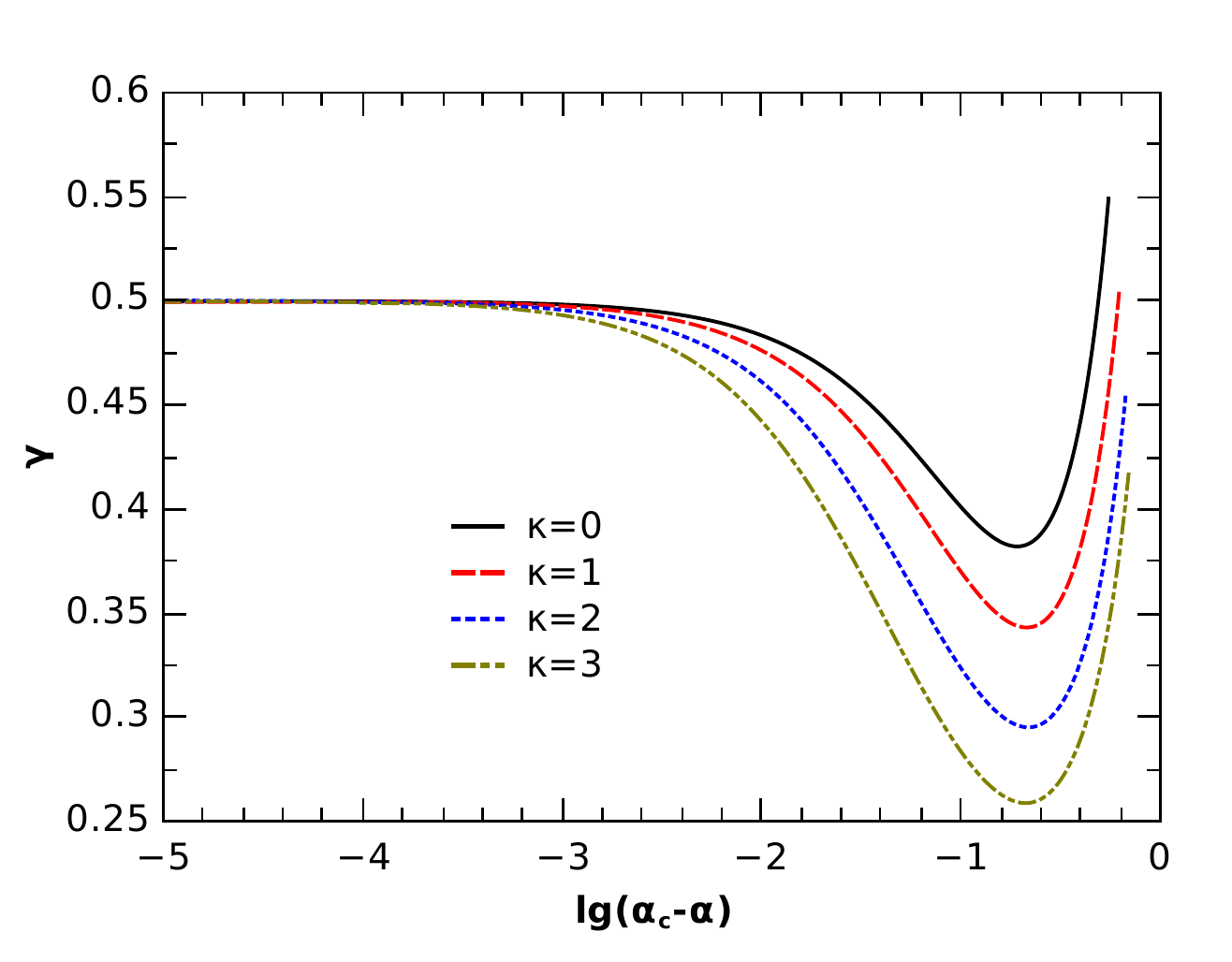}
\caption{Critical exponent of the transition 
observed in 2D three-particle systems as a function of deviation of 
$\alpha$ from its critical value $\ac$ for different strengths of 
screening.}
\label{fig3}
\end{figure}

We determine the values of the exponent $\gamma$ near the critical point by 
analyzing the slope of the above discussed log-log plot. Namely, we take the 
numerical derivative of the function 
$\lg \left( \op \right)=f \left( \lg \left(\ac-\alpha \right)\right)$. 
Calculated exactly at the critical point this derivative yields the exact
`theoretical' value of the critical exponent. However, in an experimental or
numerical investigation the precise location of the critical point may not 
be known. Thus, calculating the numerical derivative a bit away from the
critical point we are able to mimic the uncertainty and errors present in a 
realistic experimental situation. It turns out, that in all cases, $\gamma=1/2$ 
as long as $\alpha$ is close to its critical value $\ac$ (Figure \ref{fig3}). 
However, the local value of the exponent (determined as the numerical derivative) 
is very sensitive to the deviation of anisotropy parameter from $\ac$.

Figure \ref{fig3} shows the dependence of the power-law exponent 
$\gamma$ on the deviation of $\alpha$ from its critical value. The 
case of 2D cluster with $N=3$ particles and four different values of screening 
length is presented. We see, that $\gamma$ departs from the value of $1/2$ 
significantly when the deviation from $\ac$ reaches third decimal and 
attains its minimum near the first decimal. Furthermore, the exponent 
$\gamma$ attains significantly lower values far from $\ac$ in the systems with 
stronger screening. Other than that, there are no qualitative 
differences in the critical behavior of systems with different values of $\kappa$. The 
exponent $\gamma$ of the other systems with $N>3$ behaves similarly to 
the case of three particles presented here.  

A thorough analysis of transitions in a planar cluster of five particles was 
reported by Sheridan in \cite{Sheridan}. The given value of critical parameter is 
$\ac=2.96$ and the critical exponent is said to be 
$\gamma=0.39 \ne 1/2$, while we find critical anisotropy parameter to 
be $\ac=3.01$. According to the results of our calculations, a value of 
$\alpha=2.96$ corresponds to the exponent $\gamma=0.37$, which is 
close to the value reported by Sheridan et. al. Therefore, the reason of differences 
between the results, published in \cite{Sheridan} and our discoveries 
almost undeniably lies in the extreme sensitivity of $\op$ on the 
value of anisotropy parameter and distance from its critical value. Thus, the accuracy of the results 
presented in \cite{Sheridan} is probably not sufficient.

\begin{figure}[ht]
\centering
\includegraphics[width=\figa]{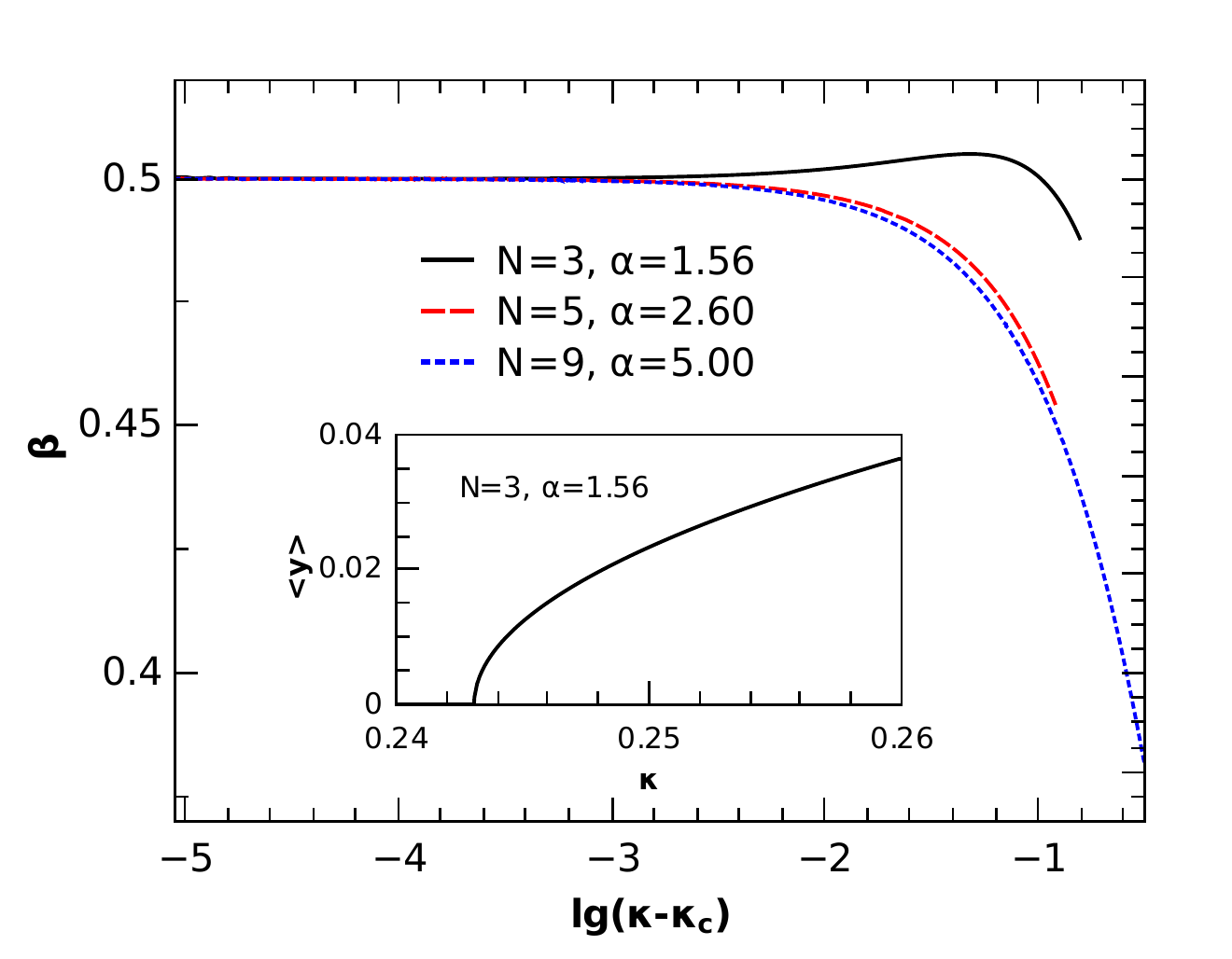}
\caption{Critical exponent $\beta$, as a function of deviation of 
$\kappa$ from its critical value $\kappa_{\rm c}$ for three 2D 
clusters. The evolution of the order parameter $\op$ close to the 
dimensional transition (inset).}
\label{figKY}
\end{figure}

As it was demonstrated in \cite{Sheridan} and shown in 
Figure \ref{fig2}, continuous transitions 
might also be induced by the variations in the screening strength 
$\kappa$. By keeping $N$ and $\alpha$ constant, we gradually change a 
value of $\kappa$ while tracking the changes parameter 
$\op$ undergoes. The dimensional transition takes place when the value of $\op$ 
suddenly drops to zero, at which point the critical value of $\kappa$ 
is obtained (see the inset of Figure \ref{figKY}). It turns out, that 
$\op$ exhibits the same power law behavior 
near the transition, i.e. $\op \propto (\kappa 
-\kappa_c)^\beta$. Figure \ref{figKY} shows the dependence of critical exponent $\beta$ 
on a logarithm of the distance from the critical value $\kappa_c$ for 
three systems. As opposed to the results presented in \cite{Sheridan}, we 
 see again, that in all cases close to the transition point 
 $\beta=1/2$. Moving away from the critical point, however, exponent 
 $\beta$ departs from the value of $1/2$ significantly. 

\subsection{ 3D $\rightarrow$ 2D transitions}

We further investigate structural transitions in three-dimensional 
Yukawa clusters with $N=4$ to $N=8$ particles and integer values of screening 
parameter up to $\kappa=3$. Increased values of the parameter $\alpha$, 
turn initially spherical structure into the oblate one; 
eventually, after the anisotropy parameter reaches its critical value 
$\ac$, dimensional phase
transition takes place and familiar two-dimensional clusters are 
formed. In three-dimensional transformations of five- and six-particle 
clusters, two 
different final states are possible, as opposed to the zigzag 
transitions in 2D, where only one linear configuration can be formed. Therefore, in 3D  
$\to$ 2D transitions, there is a distinct value of $\ac$ 
for each final configuration and we are concerned by the properties 
of phase transitions of a particular stable state. 

As the 
confinement potential well is squeezed in $y$ direction, it is handy to
label small clusters according to the arrangement of particles in 
the projection to $(xz)$ plane, in a manner similar to the state labeling 
by shell 
occupation numbers in two dimensions. Moreover, particles in three-dimensional 
anisotropic traps frequently organize themselves within the layers, 
parallel to the $(xz)$ plane. For the sake of clarity and unambiguous 
definition of the configurations, we will also use 
a list of particle numbers in distinct layers, enclosed within 
curly brackets. 

\begin{figure}[ht]
\centering
\includegraphics[width=\figa]{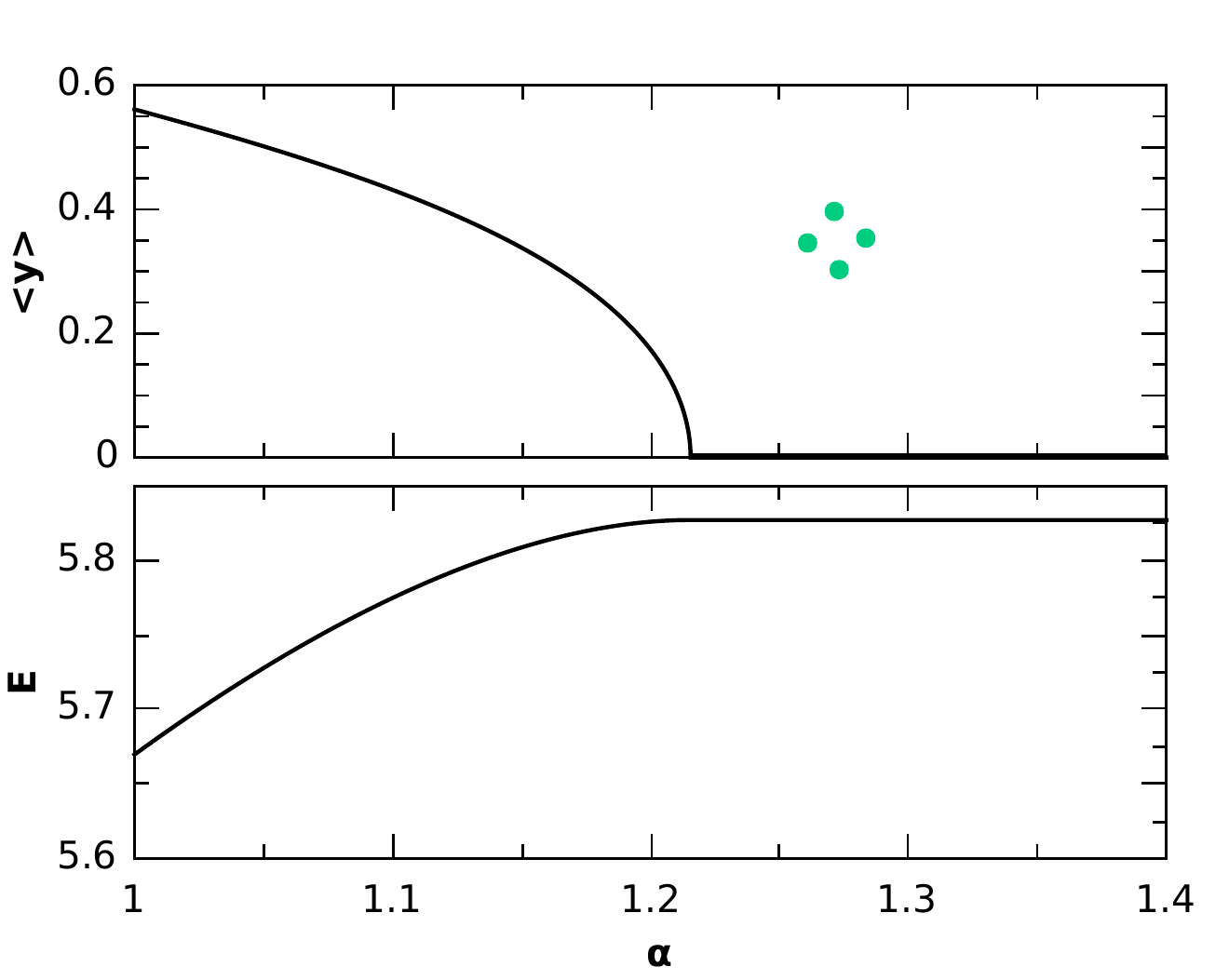}
\caption{Order parameter $\op$ and potential energy $E$ of a four-particle 
3D Coulomb cluster in the asymmetric potential trap with anisotropy 
parameter $\alpha$.}
\label{fig4}
\end{figure}

The simplest system, undergoing a non-trivial $\rm 3D \rightarrow 2D$ 
transition is the cluster composed of four particles. Not surprisingly, four particles in a 
symmetric three-dimensional trap form a regular tetrahedron, and there 
is only one possible square-shaped $(0,\, 4)$ state in two dimensions. Figure 
\ref{fig4} shows dependence of the order parameter $\op$ and potential 
energy of a system $E$ on the anisotropy parameter $\alpha$. We see, 
that $\op$ changes continuously and the transition is remarkably similar to the one in 2D case of 
$N=3$ particles. The potential energy gradually increases as the 
potential trap is flattened, until a two-dimensional structure is formed at $\ac \approx 
1.22$. As it is demonstrated in Appendix \ref{appendix}, this 
symmetric transition can be modelled analytically; the critical 
value turns out to be $\ac={({4\sqrt{2}}/(1+2\sqrt{2}) )}^{1/2}$ --- exactly the 
same as determined in our numerical modelling. Naturally, the value $\ac$ is 
sensitive to the range of the inter-particle Yukawa potential. As 
Figure \ref{fig_3D-ak} shows, critical value of the 
anisotropy parameter increases rapidly with the strength of 
screening for $\kappa<2$ and significantly slower after that, thus 
reminding of the transitions from two- to one-dimensional configurations. 

\begin{figure}[ht]
\centering
\includegraphics[width=\figa]{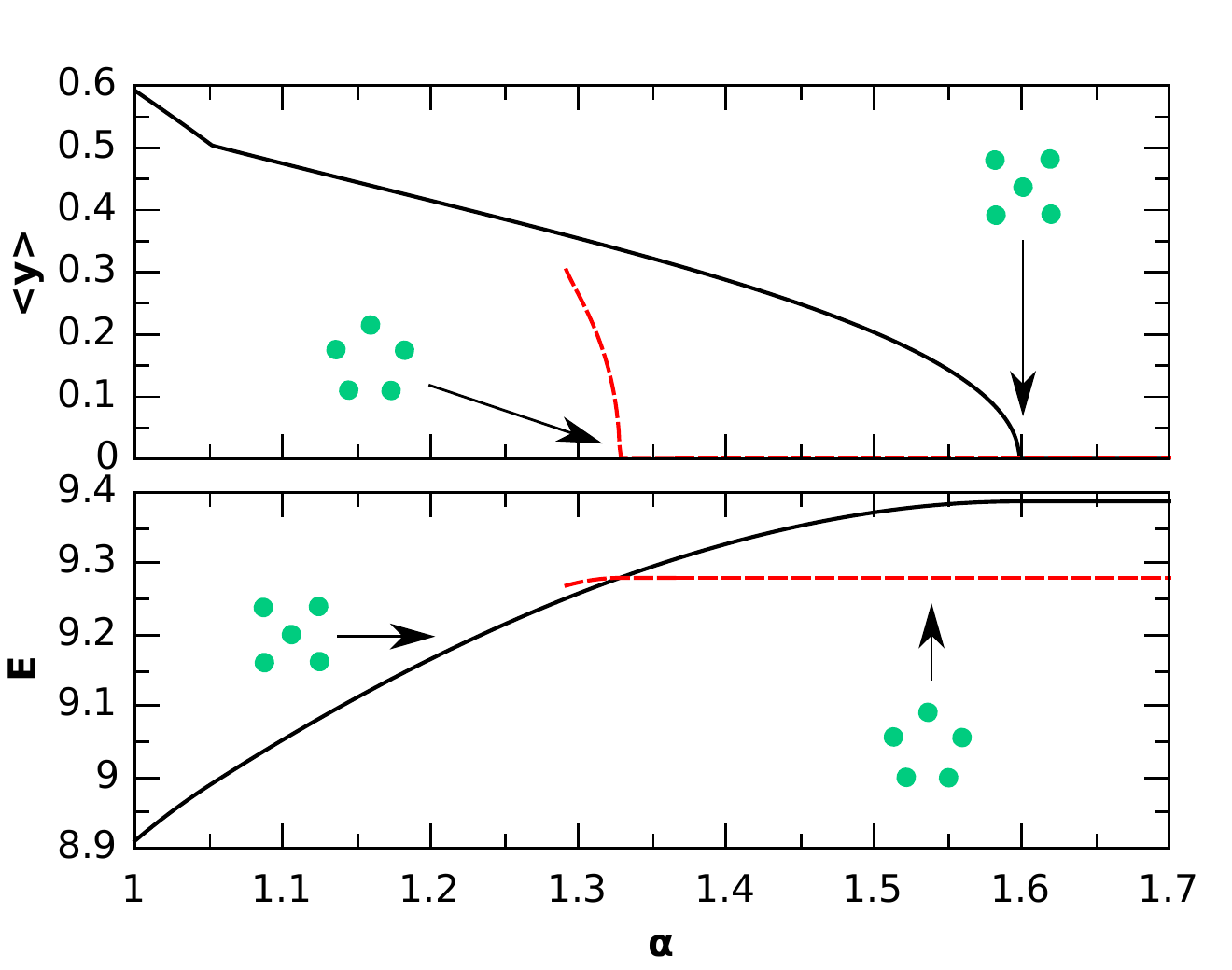}
\caption{Order parameter $\op$ and potential energy $E$ of a $5$-particle 
3D Coulomb cluster in the asymmetric potential trap with anisotropy 
parameter $\alpha$. Here and later the insets show projections of 
configurations to $(xz)$ plane.} 
\label{fig6}
\end{figure}

As it was already pointed out before, there are two competing stable 
states observed in a two-dimensional system with 
$N=5$ particles. Five particles in a spherically symmetric 3D confinement potential, 
however, can form only one stable configuration. Slightly increased 
anisotropy leads to the formation of a three-layer structure,  
with the arrangement of particles within these layers being $\left 
\{2,\,2,\,1\right\}$. As parameter $\alpha$ increases above the 
value of
$\alpha=1.05$, two layers merge forming a square and thus transforming the 
configuration into a pyramidal structure $\{ 4,\,1\}$ with 
projection $(1,\,4)_{xz}$. This 
structural transition from three-layered cluster to the pyramidal 
configuration is signified 
by the discontinuity in the derivative ${\rm d}\op/{\rm d} \alpha$ 
(Figure \ref{fig6}).

 As it is 
demonstrated in Figure \ref{fig6} for a pure Coulomb interaction, a 
second stable state appears when the anisotropy parameter reaches the 
value of $\alpha_0 \approx 1.29$. A new pentagonal state undergoes an asymmetric 
dimensional transition 
and is soon transformed into the new ground state $(0,\,5)$. The metastable  
configuration $(1,\,4)_{xz}$, on the other hand, becomes two-dimensional only at 
$\ac \approx 1.60$, through the so-called ``pyramidal'' transition mechanism. 

Both point of appearance of the second state in 
five-particle system $\alpha_0$, and  
its critical value $\ac$ depends on the type of the interaction potential 
and its screening parameter $\kappa$. As it is shown in 
Figure \ref{fig_3D-ak}, both parameters grow with the strength of screening. 
The distance between $\alpha_0$ and $\ac$, however, rapidly 
diminishes. As the screening reaches the value of $\kappa \approx 4.5$, 
two lines merge and
a new stable state appears already in its two-dimensional pentagonal form $(0,\,5)$.  

\begin{figure}[ht]
\centering
\includegraphics[width=\figa]{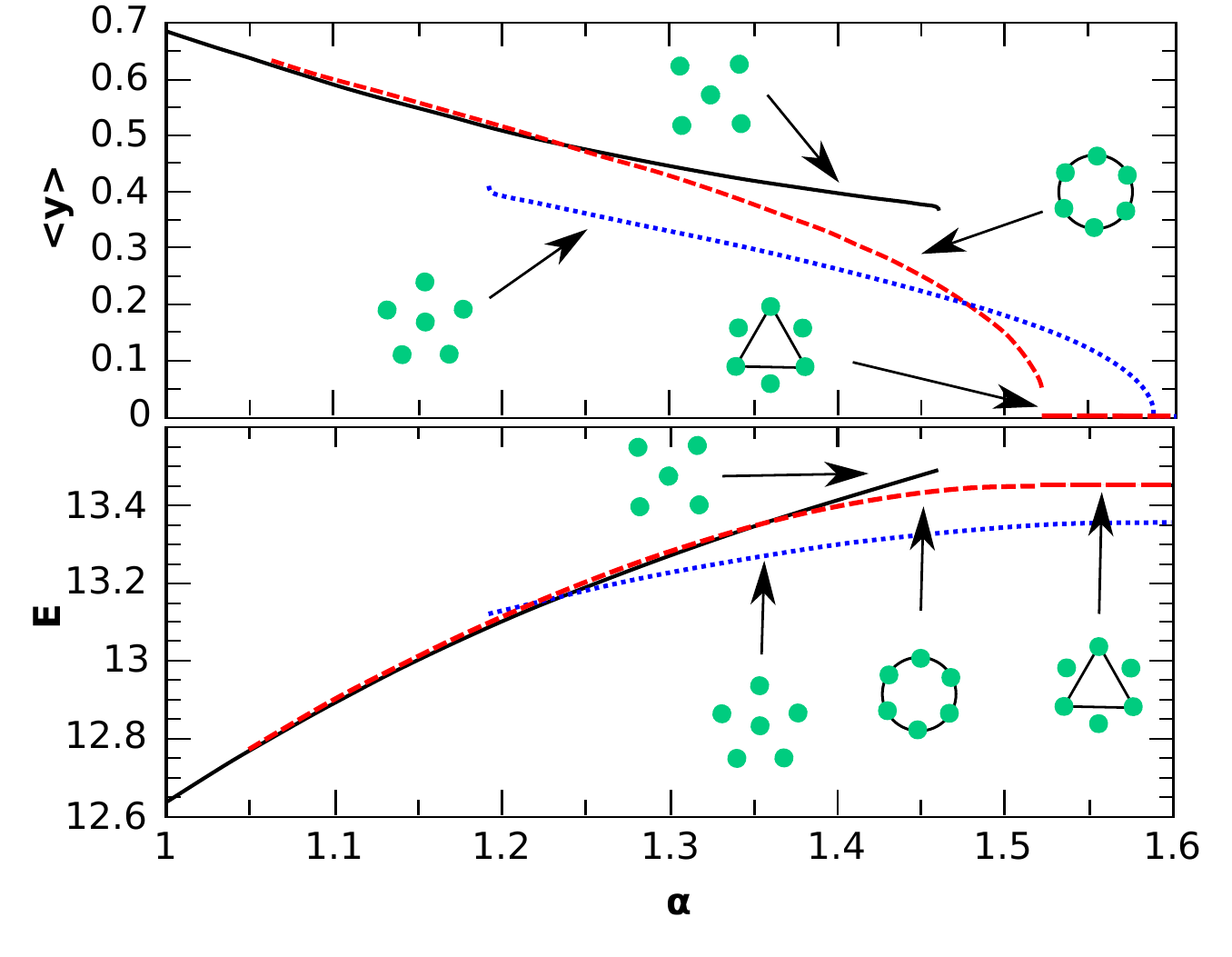}
\caption{Order parameter $\op$ and potential energy $E$ of the six-particle 
3D Coulomb cluster in the asymmetric potential trap with anisotropy 
parameter $\alpha$.}
\label{fig7}
\end{figure}

\begin{figure}[ht]
\centering
\includegraphics[width=\figa]{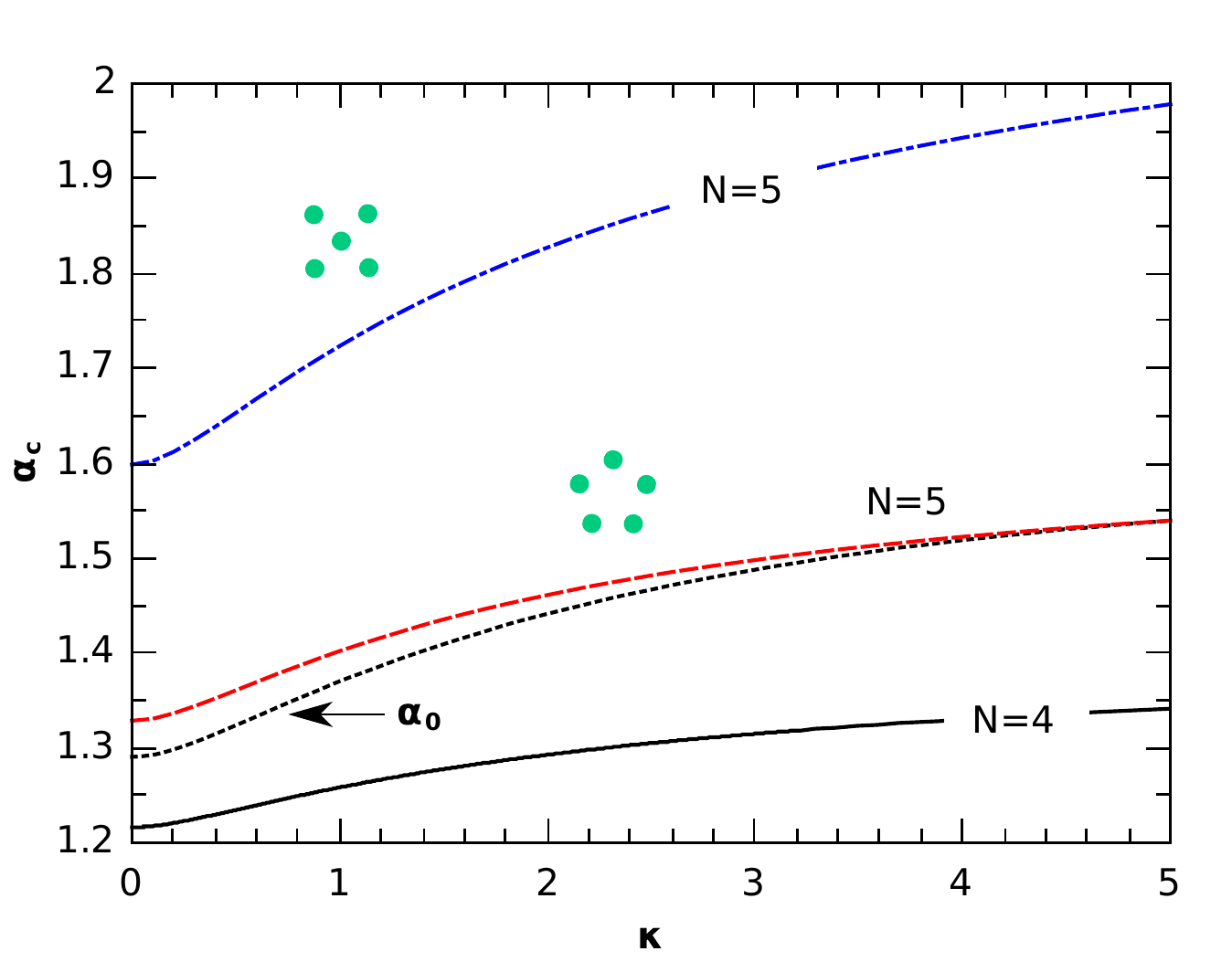}
\caption{Critical value of anisotropy parameter $\alpha_c$ as a 
function of $\kappa$ in three-dimensional configuration with $N=4$ 
particles and two 
states of $N=5$ clusters. The line marked as $\alpha_0$ 
corresponds to the appearance of the metastable state in a five-particle 
cluster.}
\label{fig_3D-ak}
\end{figure}

A pyramidal configuration might be described as a planar base, 
composed of $n=4\too 6$ particles lying parallel to the $(xz)$ plane, 
and a single particle located right above the center of the base, that 
is, configuration $\{N-1,\,1\}$,
$(1,\,N-1)_{xz}$. A pyramidal structural transition takes place, when 
the apex of the polyhedron is pushed into the base, thus becoming a 
two-dimensional configuration with only one particle in the center. A 
typical behavior of the order parameter $\op$ during such 
transitions was already discussed and is presented in figures \ref{fig6} and 
\ref{fig7}. As a matter of fact, the dimensional transitions of a 
pyramidal type can be modelled analytically and exact values of 
critical parameters $\ac$ can be found, see Appendix \ref{appendix}.

Even more stable configurations are observed in clusters with $N=6$ 
particles, as Figure \ref{fig7} shows for the Coulomb inter-particle
potential. Evolution of the system
starts with a single stable state in the symmetric 3D trap --- the octahedral 
configuration (full line in Figure \ref{fig7}). As the parameter 
$\alpha$ increases, this bipyramid is deformed  
by pushing two of its particles lying exactly on $y$-axis, 
towards each other, thus lowering the height and forming a 
configuration  $\{1,\,4,\,1\}$ with 
projection $(1,\,4)_{xz}$. $\op$ decreases slowly, until the said two particles 
start to depart form the $y$-axis near $\alpha \approx 1.46$, at which 
point a phase of rapid deformation begins. Unfortunately, right
after this happens, the stable state disappears. The same scenario of bipyramidal 
deformation also applies to the larger clusters, e.g. $N=7,\;8,\;9$, 
and has a specific, well recognizable shape of its $\op=f(\alpha)$ 
curve, with the segments of slow and rapid changes.

A new metastable state emerges near $\alpha \approx 1.06$: the 
particles lie on the six vertices of two
parallel equilateral triangles, centered precisely on $y$-axis and rotated 
by $\pi/3$ with respect to each other, that is, state $\{3,\,3\}$ and 
$(0,\,6)_{xz}$. These triangular layers are pushed 
towards each other by deformations of the confinement, however, they 
fail to ever become a 
truly two-dimensional configuration. Instead, as it is demonstrated in 
Figure \ref{fig7}, the configuration ceases to exist at $\alpha \approx 1.52$ 
where the r.m.s. value of $y$ coordinate is still $\op\approx 0.05>0$.
However, right before 
the disappearance, a new similar purely two-dimensional state shows 
up. The new planar configuration is 
composed of six particles lying on the vertices of two
triangles of slightly different sizes (see Figure 
\ref{fig7}). Therefore, in a brief range of 
$\alpha$ values these two states exist simultaneously and there is no 
continuous transition between them. Finally, a pyramidal 
configuration $\{5,\,1\}$ appears near $\alpha 
\approx 1.19$ and undergoes the usual pyramidal dimensional transition at $\ac \approx 
1.59$, the value predicted by our analytical model (appendix 
\ref{appendix}).

\begin{figure}[ht]
\centering
\includegraphics[width=\figa]{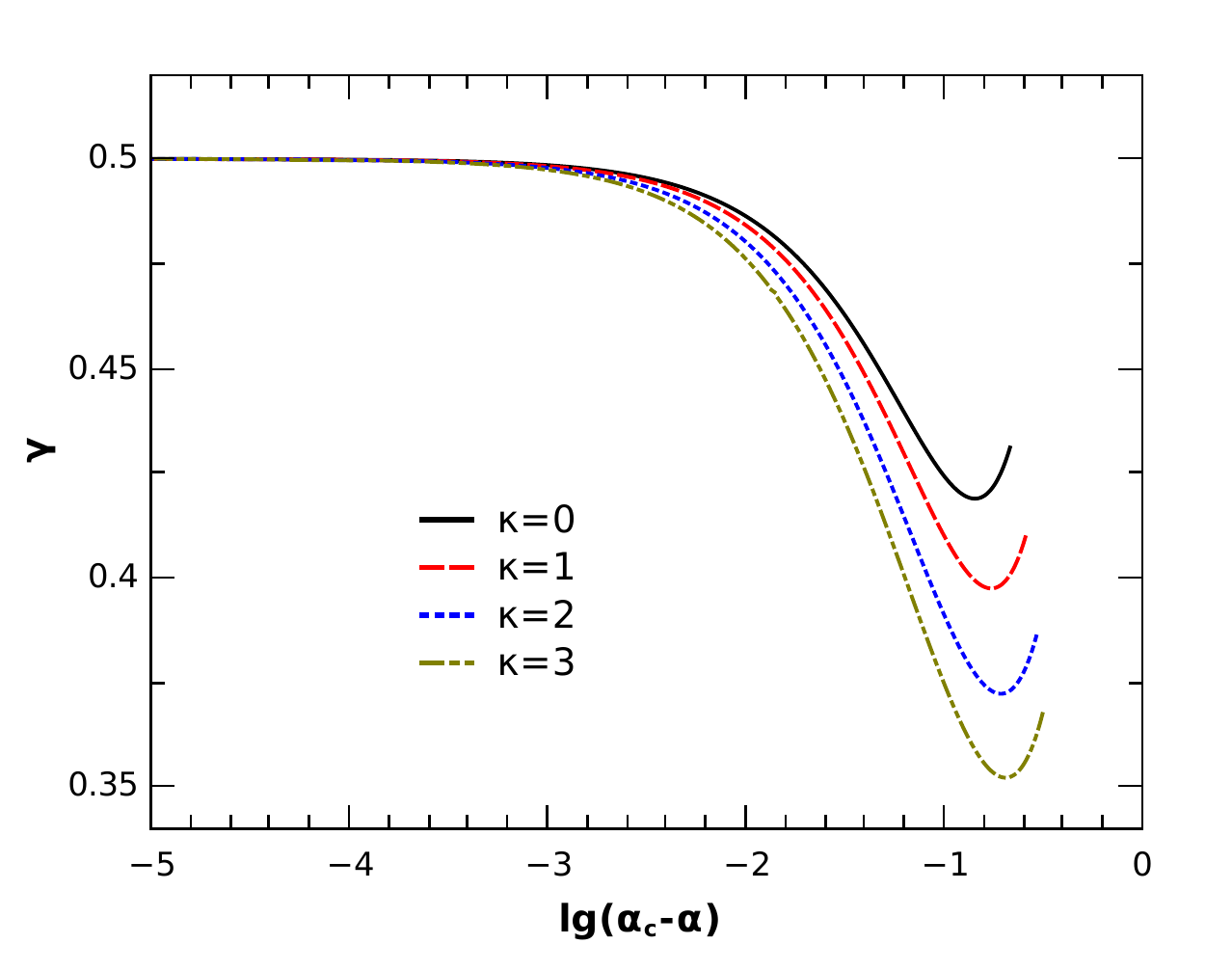}
\caption{Critical exponent $\gamma$ for a dimensional transition, 
observed in 3D four-particle systems as a function of deviation of 
$\alpha$ from its critical value $\ac$ for different strengths of 
screening.}
\label{fig5}
\end{figure}

Close to the critical point of continuous transitions from three- to 
two-dimensional systems, a 
power-law behavior of the order 
parameter is detected once again, i.e. $\op \propto (\ac 
-\alpha)^\gamma$. In the same manner as 
in 2D case, Figure \ref{fig5} shows the dependence of the power-law exponent 
$\gamma$ on the logarithm of  $\ac-\alpha$. It turns out, that in a close vicinity of transition 
point, the critical exponent $\gamma=1/2$ does not depend on the 
screening strength $\kappa$. Deviations from this value occur when the departure of 
$\alpha$ from its critical value reaches third decimal. Just as in the 
two-dimensional case, the value of a critical exponent is lower for systems with 
stronger inter-particle potential screening, and drops as low as 
$\gamma \approx 0.35$ for $\kappa=3$. Essentially the same behavior 
of the exponent $\gamma$ is observed in larger three-dimensional systems, 
where dimensional transitions take place, be it 
pyramidal transitions or transformations of any other type.

\begin{figure}[ht]
\centering
\includegraphics[width=\figa]{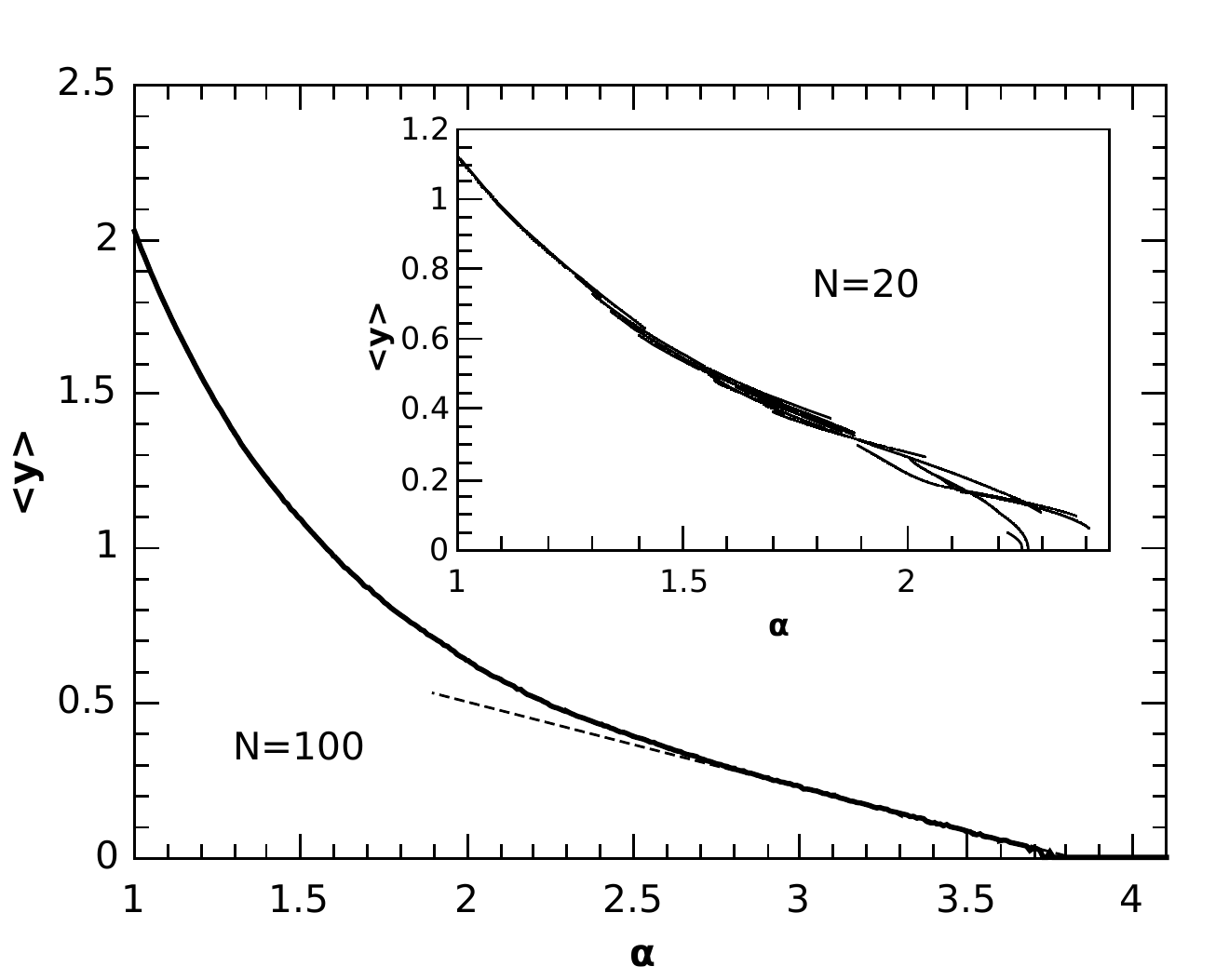}
\caption{Order parameter $\op$ of large Coulomb systems with 
$N=100$ and $N=20$ (inset) 
particles as a function of the anisotropy $\alpha$.}
\label{fig10}
\end{figure}

As the number of particles grows, more and more stable states emerge 
and, as a consequence, $\op=f(\alpha)$ graphs become convoluted and 
somewhat difficult 
to study. An illustrative example is given in the 
inset of
Figure \ref{fig10}, where the behavior of the order parameter in 
stable states of
20-particle 
Coulomb system is presented. 
We can still see a few distinct continuous phase transitions in the 
vicinity of $\alpha \approx 2.26$, however different lines become 
hardly distinguishable 
at the lower values of $\alpha$. In very large systems, the 
values of $\op$ for all metastable states lie virtually on the same 
line, as it is illustrated in Figure \ref{fig10} with a 100-particle 
cluster and $\kappa=0$. 

It might be worth discussing the structural evolution of Yukawa 
clusters confined by traps 
with a prolate equipotential surface. In our model, this effect is achieved by 
lowering the value of anisotropy parameter $\alpha$ towards 
zero. In that way, elongated clusters are formed, with low potential energies 
and high values of $\op$. Consequently, in order to study this type of 
structural transitions, a new order parameter must be 
defined. We choose to rely on the root mean square of the distance from $y$-axis:
\begin{equation}
\ro = \left(\frac{1}{N} \sum_{i=1}^N 
\left ( x_i^2+z_i^2 \right ) \right )^{1/2}.
\end{equation}

\begin{figure}[ht]
\centering
\includegraphics[width=\figa]{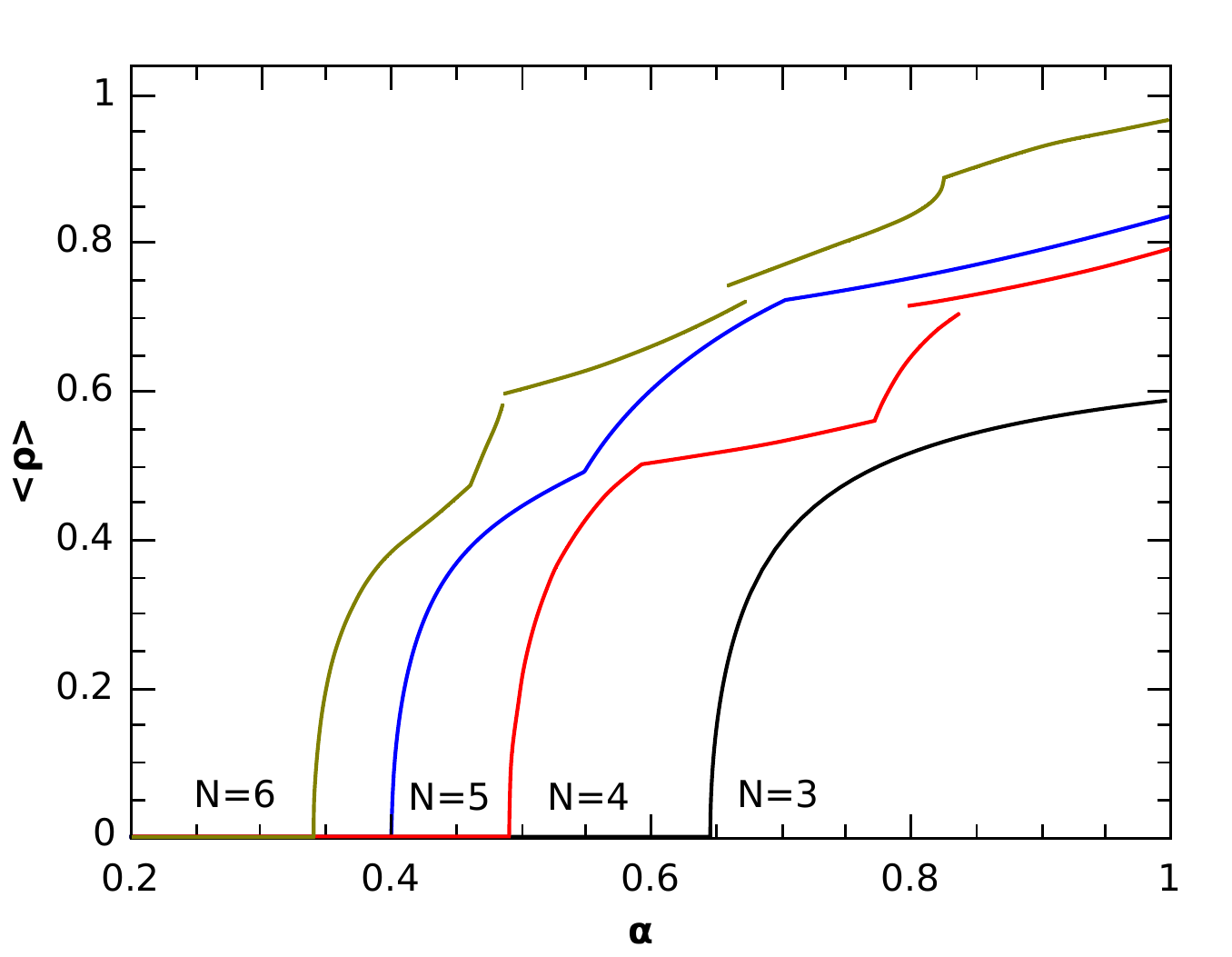}
\caption{Order parameter $\ro$ as a function of the 
anisotropy $\alpha<1$ for Coulomb clusters with $N=3\too 6$ particles.}
\label{fig11a}
\end{figure}

As Figure \ref{fig11a} shows for $N=3 \too 6$, dependencies of the order parameter $\ro$ 
on the anisotropy $\alpha$ are not smooth and in some cases feature discontinuities. We 
conclude, that in fact there is no direct transition 
from three- to one-dimensional configurations. Instead, the system is first 
transformed into the elongated 2D zigzag pattern, and only later, $\rm 2D 
\rightarrow 1D$ structural transition takes place. With that in mind, 
there is no surprise, that the values of critical parameters $\ac$ 
found by lowering $\alpha$ are the exact
inverses of those, determined in subsection \ref{sec:2D1D}. 
  
Large three-dimensional clusters in prolate traps become 
one-dimensional through the mechanism, which seems to be universal for 
all values of $N$ used in our modelling. At first, the system is squeezed and elongated 
until the particles arrange themselves into the shape of double-helix. As $\alpha$ is 
lowered further, a number of helical turns decreases, until the helix 
unwinds and cluster becomes a two-dimensional zigzag configuration. 2D system then 
undergoes the usual zigzag transition with the power-law behavior 
near the critical point.

\section{Conclusion}
\label{Sum}

Confined Yukawa clusters are among the physical systems, where simple 
interparticle interactions lead to the emergence of complicated patterns and 
spontaneous ordering. In this article, we present our findings in the numerical 
and analytical studies of two- and three-dimensional clusters confined by 
asymmetric parabolic traps. 

We confirm, that dimensional transitions from oblate three- to 
two-dimensional systems as well as from planar to linear 
configurations can be induced by changes 
in the anisotropy of the confinement $\alpha$, and screening 
strength $\kappa$. On the other hand, 
there are no direct transitions from three- to one-dimensional systems 
in prolate harmonic traps; two-stage 
transformations take place instead.

A critical value of the anisotropy parameter in general grows with the screening strength 
$\kappa$. The growth is steepest for small values of $\kappa$ and 
almost saturates for the large ones.
In a close vicinity of dimensional phase transition, the order parameter $\op$ 
exhibits power-law dependence on a control parameter, be it 
$\alpha$ or $\kappa$. 

In all cases studied here, the critical exponent is found to be 
universal and equal to $1/2$, which is consistent with the general theory of 
second order phase transitions. However, a value of the power-law exponent turns 
out to be very sensitive to the deviations of a control parameter from 
its critical value. Far from the critical point, the exponent attains lower values in
systems with stronger screening and shorter range of the inter-particle 
interaction.

\appendix
\section{Analytical values of $\ac$}
\label{appendix}

In a few simplest cases of high symmetry, when total energy of a 
cluster after a 
 transition depends on a single 
generalized coordinate, values of a critical anisotropy parameter $\ac$ can be 
identified analytically. These are the transitions in 3-particle 2D 
system, 4-particle 3D cluster and all of the pyramidal transitions.

Consider the $N$-particle three-dimensional system undergoing a 
pyramidal $\rm 3D \rightarrow 2D$ dimensional transition. Immediately 
after the transition, a planar cluster consists of $n=N-1$ particles, 
positioned on the circumference of a circle with radius $R$, and a 
single particle in the center of confinement. The 2D cluster lies 
in $(xz)$ plane. The total potential energy of the system can be 
expressed as
\begin{equation}
U_0(R) = \frac{1}{2}nR^2 + \frac{n}{R}f(n) + \frac{n}{R}.
\end{equation}
A second term here represents the Coulomb interaction energy of $n$ particles 
positioned on the
circle with radius $R$, thus forming a regular polygon. The function $f(n)$ 
 depends only on the number of particles and is 
\begin{equation}
f(n)= 
\begin{cases}
\displaystyle
\frac{1}{4}+\frac{1}{2}\sum_{m=1}^{n/2-1} {\sin^{-1}\left(m\pi/n\right)} & \text{if $n$ is 
even,} \\
\\
\displaystyle
\frac{1}{2}\sum_{m=1}^{(n-1)/2} {\sin^{-1}\left(m\pi/n\right)}& \text{if $n$ is 
odd}.
\end{cases}
\end{equation} 
By setting $\partial U_0/\partial R=0$ we find the equilibrium radius 
to be simply
\begin{equation}
\label{eq:radius}
R = \left(1+f(n)\right)^{1/3}.
\end{equation}

We further perturb the system by shifting the base by the small distance 
of $\dy$, and, keeping 
the center of mass at the center of the confinement, an apex by the 
distance of $n \delta y$ in an opposite direction. Now, the total energy 
of the system is
\begin{equation}
\begin{split}
U_{\rm p} & =\frac{1}{2}n(R^2+\alpha^2\dy^2)+\frac{1}{2}\alpha^2 
{(n\dy)}^2\\
&+\frac{n}{R}f(n)+\frac{n}{\sqrt{(R^2+(n+1)^2\dy^2}}.
\end{split}
\end{equation}
By expanding the right hand side in powers of $\dy$ up to the quadratic 
term, we get
\begin{equation}
U_{\rm p}\approx U_0+\frac{1}{2}n\alpha^2{\dy}^2(1+n) - \frac{n(n+1)^2\dy^2}{2R^3}.
\end{equation}
By requiring that
\begin{equation}
\delta U = \frac{1}{2}n\ac^2\dy^2(1+n) - \frac{n(n+1)^2\dy^2}{2R^3}=0
\end{equation}
for any small $\dy$, and making use of equation \ref{eq:radius}, we 
find the critical value of anisotropy parameter to be
\begin{equation}
\label{eq:ac}
\ac = \sqrt{\frac{N}{f(N-1)+1}}.
\end{equation}

\begin{table}[h]
\caption{Values of the function $f$ and critical parameter $\ac$ of 3D 
Coulomb systems undergoing dimensional transition of the pyramidal type.}
\label{tab:ac}
\begin{tabular}{|c|c|c|}
\hline
$N$ & $f(N-1)$ & $\ac$\\
\hline
\hline
$5$ & $\displaystyle \frac{1}{4}+\frac{\sqrt 2}{2}$& $ \displaystyle 2 \sqrt{\frac{5}{5+2 
\sqrt{2}}} \approx 1.5983715 $ \\
\hline
$6$ & $ \displaystyle \sqrt{1+\frac{2}{\sqrt 5}}$ & $ \displaystyle \sqrt{3 \sqrt{5+2 
\sqrt 5}-\sqrt 5} \approx 1.5889766$ \\
\hline
$7$ & $ \displaystyle \frac{5}{4} + \frac{1}{\sqrt 3}$& $ 
\displaystyle 2 \sqrt{\frac{21}{27 + 4 
\sqrt{3}}} \approx 1.5734727$\\
\hline
$8$ & 2.3047649 & 1.5558750\\
\hline
\end{tabular}
\end{table}
Values of the function $f(n)$ and corresponding critical parameters 
of pyramidal transitions are collected in table \ref{tab:ac} for all 
3D clusters with this type of structural transformation. We see, that 
in general, $\ac$ slightly decreases with $N$ in the range of $5\too 
8$ particles.

In the two-dimensional case of $N=3$ particles, the system forms a 
triangular cluster. During the structural transition, one of its 
particles is pushed in-between the others, thus forming a linear 
structure. This case is basically a generalization 
of pyramidal transitions to the two-dimensions. Therefore, equation 
\ref{eq:ac} is still valid and we find critical anisotropy to be 
$\ac=\sqrt{12/5}$, which is exactly the value observed in our 
numerical modelling.

A slightly different mechanism of transformation is observed in $\rm 3D \rightarrow 2D$ 
transition of highly symmetrical $N=4$ particle cluster. The final 
configuration is a 2D square, with potential energy 
\begin{equation}
\label{eq:U4}
U_0(R) = \frac{1}{2}NR^2 + \frac{N}{R}f(N).
\end{equation}
Minimization of \ref{eq:U4} by solving $\partial U_0 / \partial R=0$ in 
turn gives $R=f(N)^{1/3}$. 
Right before the transition, two particles sharing a common diagonal 
in the final square are elevated by the distance of $\dy$ above $(xz)$ 
plane, while other two are located at the same distance below it; ergo the energy of the 
perturbed system
\begin{equation}
\begin{split}
U_{\rm p} &= \frac{1}{2}N(R^2+\alpha^2 \dy^2) + \frac{1}{4}\frac{N}{R} 
+ \frac{N}{\sqrt{ 2R^2 + 4\dy^2 }} \approx \\
&\approx U_0 + \frac{1}{2}N\alpha^2\dy^2 - \frac{N \sqrt 2 
\dy^2}{2R^3}.
\end{split}
\end{equation}
By setting $\delta U=U_{\rm p}-U_0=0$, we again get
\begin{equation}
\ac = \sqrt{\frac {\sqrt 2}{f(4)}} = 
\sqrt{\frac{4\sqrt{2}}{1+2\sqrt{2}}}\approx 1.2155625.
\end{equation}
This value is exactly the same as found by our numerical procedure.


\end{document}